# Percolation-Driven Magnetotransport due to Structural and Microstructural Evolution in Ultrathin Si/Fe Bilayers

S. S. Das[1] and M. Senthil Kumar[2]

[1]Institute of Industrial Science, The University of Tokyo, 4-6-1 Komaba, Meguro-ku, Tokyo 153-8505, Japan

[2]Department of Physics, Indian Institute of Technology Bombay, Mumbai 400 076, India

E-mail: senthil@iitb.ac.in, ssdas@iis.u-tokyo.ac.jp

**Abstract:** The anomalous Hall effect (AHE) in magnetic nanofilms is highly sensitive to the microstructural and magnetic homogeneity. However, the evolution of the microstructure and morphology near the percolation threshold, and its connection to the resulting magnetic and magnetotransport behavior in low-dimensional magnetic heterostructures, remain poorly understood. In this study, we present a comprehensive analysis of the evolution of the structural, microstructural, and magnetotransport properties of Si/Fe bilayers by varying the Fe layer thickness in the range of 10 -200 Å. X-ray diffraction (XRD) and high-resolution transmission electron microscopy (HRTEM) analyses reveal a percolation-driven transition from a continuous metallic film to percolative network structure of grains when $t_{Fe}$ decreases below 30 Å. Magnetic measurements confirm this transition as well, showing a sharp decrease in saturation magnetization ($M_s$) below 30 Å. Transport measurements involving longitudinal resistivity ($\rho$), and the anomalous Hall resistivity ($\rho_{h,s}^A$) show clear divergence near the percolation threshold. The purely electronic conduction channels ($\rho$) evolve more gradually as compared to the combined electronic and magnetic ones ($\rho_{h,s}^A$). The percolative analysis of the structural, magnetic, and magnetotransport data yields a critical exponent in the range of 0.78 to 1.16, consistent with that of two-dimensional disordered systems. The AHE scaling relation between the $\rho_{h,s}^A$ and $\rho$ reveals a crossover of the AHE mechanism from a mixed intrinsic/side-jump contribution with a minor skew scattering component (n ~ 1.42) in the thick, low-resistive samples ($t_{Fe}$ > 30 Å) to a skew-scattering-dominant mechanism (n = 0.62) in the high-resistive films ($t_{Fe}$ ≤ 30 Å). This crossover coincides with the onset of structural and magnetic connectivity between the grains. Furthermore, these findings underscore the interlink between microstructure, morphology, magnetism, and Hall transport under a percolation framework, offering opportunities to tune the AHE response in low-dimensional systems for spintronic applications.

# 1. Introduction

In recent years, owing to their applications in the field of spintronics, magnetic sensing, and memory devices, ultrathin ferromagnetic films and their heterostructures have attracted significant research attention [1–5]. With the rapid evolution of semiconductor technology, the metal-semiconductor hybrid structures involving Si and Fe, have garnered significant focus by offering a unique platform to investigate the ferromagnetic-semiconductor interface and its effect on controlling the spintronic properties [6–11]. Furthermore, the structural, microstructural, magnetic, and magnetotransport properties of the system can be modulated through effective tuning of the interfacial layer thickness. When the thickness of the ferromagnetic layer reaches the atomic scale, new phenomena such as quantum confinement, surface-interface scattering, and finite-size effects emerge, causing large deviations from bulk behavior and giving rise to novel transport mechanisms, thereby attracting substantial research interest.

The functional properties of thin films, such as their electronic, magnetic, and magnetotransport responses, depend strongly on their morphology, microstructure, interface quality, and the degree of percolative connectivity. Following the Volmer-Weber growth mode [12–14], film growth on a substrate begins with the formation of small, isolated islands separated by a finite distance. As the thickness increases, these islands coalesce to form isolated clusters separated by a small distance. As the thickness increases further, these clusters elongate and subsequently get interconnected to form a network structure. Above a certain thickness $t_c$, these cluster networks grow further, forming an "infinite cluster" (a single, dominant, spanning network) that covers the entire substrate. As $t$ increases further ($t \gg t_c$), the area fraction of the infinite cluster increases rapidly, forming a continuous film [15]. This percolation transition not only affects the film morphology, but also the electronic, magnetic, and magnetotransport properties of the material, which include anomalous Hall effect (AHE) and anisotropic magnetoresistance (AMR). In the percolating regime, the conduction mechanism is controlled by electron scattering at the grain boundaries, surfaces, and interfaces, resulting in a large enhancement in the electronic resistivity of the material. The AHE [3], which has its origin in the spin-dependent scattering of charge carriers at the localized magnetic centers, can undergo enhancement in the percolating regime. Furthermore, the non-trivial Hall signals are found to be sensitive to nanoscale connectivity and structural coherence [3,16,17]. Recently, Gerber et al. demonstrated that the magnetic transition from the ferromagnetic phase to the superparamagnetic phase in CoPd alloys follows the classical percolation model, with the transition occurring at a threshold higher than the conductivity threshold [18]. Since AHE is the electrical replica of magnetization, it can provide information about magnetic transitions in thin film samples [18–20]. Just as the electrical conductivity gives information about structural continuity in the thin film samples, the AHE, which depends on the magnetization of the material, provides information about the magnetic inhomogeneity in nanostructured systems. AHE being dependent on the scattering event can be scaled with $\rho$ of the material. Previous studies have shown that the AHE in thin films can be strongly influenced by size effects, interfacial roughness, spin-orbit coupling, and the degree of percolation [3,21]. However, the detailed correlation between structural transitions and magnetotransport properties in Si/Fe bilayers, especially in the ultrathin film limit where the film morphology transitions from isolated islands to interconnected networks and finally to continuous layers, remains underexplored.

In this study, we present a comprehensive analysis of percolation-driven transitions in the structural, microstructural, magnetic, and magnetotransport properties of ultrathin Si/Fe bilayers with varying Fe layer thicknesses. By precisely adjusting the growth parameters near the percolation threshold, we observed significant changes in the materials' magnetic and transport behavior. Our findings emphasize the importance of percolation-driven transitions in altering the AHE mechanism in Si/Fe bilayers and provide insights into engineering magnetic nanostructures with tailored functionalities for integration into spintronic devices.

## 2. Experimental Details

A series of Si/Fe bilayers was prepared by DC magnetron sputtering at ambient temperature on three different substrates: glass, silicon (111), and carbon-coated copper grids. The glass and Si substrates were used to examine the possible substrate effects on film growth and percolation behavior. However, as will be seen later in the results section that there is no difference in the behaviour of the films on these substrates was observed. Carbon-coated Cu grids were used for preparing selected samples intended for transmission electron microscopy (TEM) analysis because thick substrates of glass or Si cannot be used in transmission mode of the experiment. All films were deposited in the sequence substrate/Si(50 Å)/Fe($t_{Fe}$), where the Si underlayer provides a smooth, diffusion-blocking template that suppresses intermixing and promotes uniform Fe nucleation.

The thickness of the Fe layer ($t_{Fe}$) was varied from 10 to 200 Å, while that of the Si layer was kept fixed at 50 Å, resulting in a nominal composition of the bilayers as Si(50 Å)/Fe($t_{Fe}$). The base vacuum achieved prior to the deposition was $2 \times 10^{-6}$ mbar, and sputtering was carried out in Ar at a pressure of $4 \times 10^{-3}$ mbar. Both the Fe and Si targets were sputtered at a DC power of 40 W. The deposition conditions were kept the same as in reference [7]. The prepared samples were subjected to various characterizations to analyze their structural, microstructural, magnetic, and magnetotransport properties. Structural properties were investigated using X-ray diffractometry (XRD) with a PANalytical X'Pert Pro X-ray diffractometer with Cu$K_\alpha$ radiation of wavelength $\lambda = 1.54184$ Å. The microstructure of the samples was examined through imaging and diffraction using a JEOL JEM 2100F high-resolution transmission electron microscope (HRTEM). TEM imaging and selected-area electron diffraction (SAED) were performed using the same instrument operated at 200 kV, with samples deposited directly onto carbon-coated Cu grids for plan-view analysis. Magnetic characterizations were performed in both the in-plane and perpendicular directions at 300 K by using a vibrating sample magnetometer (VSM), an attachment of a Physical Property Measurement System (PPMS) from Quantum Design, Inc., with a magnetic field up to 50 kOe. Hall effect measurements were performed on the samples at 300 K using the four-point probe technique in external magnetic fields varying up to 27 kOe, perpendicular to the film plane.

## 3. Results and discussion

In this section, we discuss the results that reveal how the structural and microstructural evolution in ultrathin Si/Fe bilayers affects their electronic, magnetic, and magnetotransport properties. First, we examine the structural and microstructural changes in the sample as it undergoes transformation from an island-type morphology to a percolating network structure and finally to a continuous film. Then, we correlate these changes with the magnetic and magnetotransport behavior of the sample.

In order to analyze the data further, we adopt the percolation model. Fig.1 shows the schematic of morphological transition in a thin film (upper panel) and its corresponding changes in the resistivity ($\rho$) and Hall resistivity ($\rho_h$) at different stages of its growth. In the percolative framework, a fraction $p$ of the available sites is occupied by the conducting clusters. Below a critical fraction $p_c$, at the initial growth stage, the film morphology is an isolated island type with a well-defined separation distance between them. With the increase of growth, above $p_c$, an infinite cluster spanning over the entire substrate appears and the dc conductivity of the sample increases nonlinearly following a power law $\sigma \propto (p - p_c)^t$ [22–24], where $p$ is the occupied conducting fraction, $p_c$ is the percolation threshold and $t$ is the conductivity (transport) exponent which takes value 1.3 in 2D and 2 in 3D. For a film (Fe, here) growth at fixed morphology, the conducting fraction $p$ can be replaced with the effective Fe layer thickness $t_{Fe}$, resulting $\sigma \propto (t_{Fe} - t_{Fe,c})^\mu$ where µ is the transport exponent. We applied this percolation law to describe the grain growth, ρ, and Hall resistivity $\rho_h$ behaviour near the percolative regime.

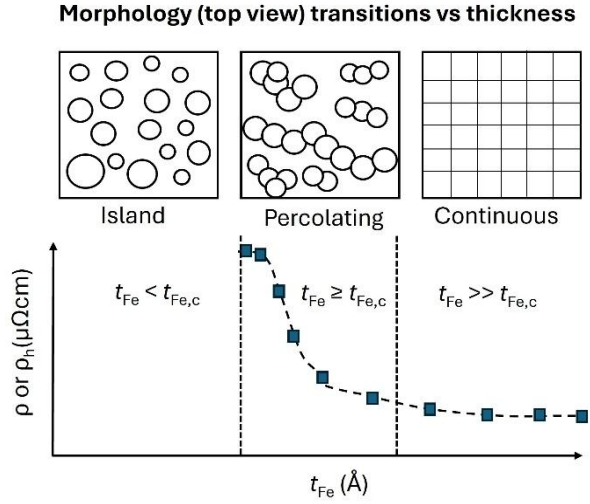

**Fig. 1** Schematic showing the morphological transition in a thin Fe film with thickness. The top panel illustrates the morphology of the film as it progresses through three stages: island ($t_{Fe} < t_{Fe,c}$), percolating ($t_{Fe} \geq t_{Fe,c}$), and continuous ($t_{Fe} \gg t_{Fe,c}$). The bottom panel illustrates the behavior of ρ and $\rho_h$ during this transition.

**(A) Structural properties**

We investigated the structural properties of the Si(50 Å)/Fe($t_{Fe}$) bilayer samples using XRD. Fig. 2 shows the XRD data of the bilayer samples with $t_{Fe}$ varying in the range from 10 to 200 Å. These samples were scanned within an angular range of 20° to 90°. Only one distinct XRD peak was observed at around 2θ = 44.5°, suggesting the presence of nanocrystalline bcc Fe phases. XRD scans within the 35° – 55° angular range (Fig. 1a) reveal the evolution of the observed Fe(110) peak. The vertical dashed line indicates the position of the standard bulk value for the Fe(110) reflection. For the samples with $t_{Fe} \leq 50$ Å, the XRD intensity is too weak to be noticeable in the plotted spectra; however, the Fe(110) peak is still visible upon separate inspection in an expanded scale, indicating the presence of Fe nanocrystallites near the XRD detection limit. As $t_{Fe}$ increases, the Fe (110) peak becomes sharper (reduced full width at half maximum, FWHM), consistent with grain growth and improved crystallinity. No peaks were observed from the Si layer, which suggests its amorphous nature.

The average grain size was estimated from the Fe(110) XRD peak, using the Scherrer formula [25] $\delta = (0.9 \cdot \lambda)/(\beta \cdot \cos\theta)$, where $\lambda$ is the X-ray wavelength and $\beta$ is the FWHM of the Fe(110) peak (in radians), and the $\delta$ values are plotted in Fig. 2(b) as a function of $t_{Fe}$. The open squares represent the data points, and the solid line is a fit using the equivalent percolation model [22,24] for grain growth $\delta = \delta_0 + A \cdot (t_{Fe} - t_{Fe,c})^\mu$. Here, $\delta_0$ refers to the thickness-independent term dominant below the critical thickness $t_{Fe,c}$, A is the scaling factor, and μ is the scaling exponent. The fit yields $\delta_0 = 15.53$ Å, $t_{Fe,c} = 49.77$ Å, and μ = 0.205. This suggests that below $t_{Fe,c} = 49.77$ Å, the grain size δ remains almost independent ($\delta_0 = 15.53$ Å) of $t_{Fe}$. Above $t_{Fe,c}$, however, it follows a percolation power law $A \cdot (t_{Fe} - t_{Fe,c})^\mu$, with the scaling exponent μ = 0.20, which suggests a more gradual grain coalescence process. This value is smaller than the two-dimensional exponent value of 1.3 [22], indicating that the grain growth is limited by island connectivity and limited adatom movement at the growth temperature. This microstructural evolution is crucial for controlling the magnetic and magnetotransport behavior of the samples, as discussed in the subsequent sections.

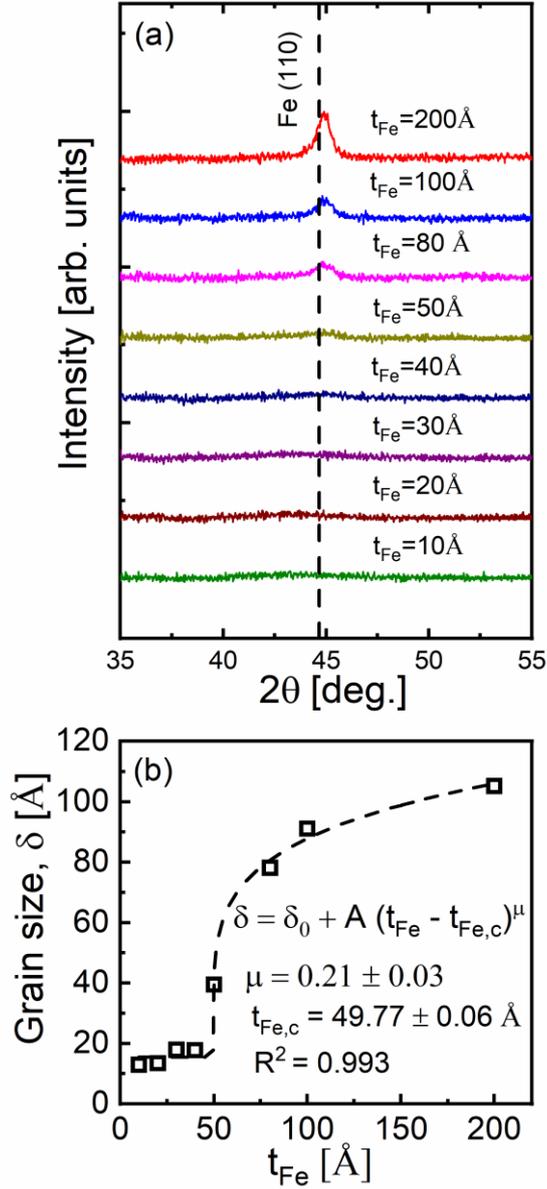

**Fig. 2** (a) XRD patterns of the Si(50 Å)/Fe($t_{Fe}$) bilayers. The vertical dashed line shows the position of the standard bulk value of the Fe(110) reflection. (b) The $t_{Fe}$ dependence of $\delta$ obtained from the Fe(110) peak. Here, the open squares are the data points, and the dashed line is a fit using the equation $\delta = \delta_0 + A \cdot (t_{Fe} - t_{Fe,c})^\mu$, where $\delta_0 \sim 15.5$ Å is independent of $t_{Fe}$ below $t_{Fe,c}$ (~ 49.77Å), and above $t_{Fe,c}$, it follows a power law with exponent $\mu = 0.21 \pm 0.03$.

**(B) Microstructural properties**

To gain additional insight into the structural aspects of the Si/Fe bilayers, we have investigated their microstructure using HRTEM imaging and SAED pattern analysis. Representative HRTEM images of the Si(50 Å)/Fe(10 Å) and Si(50 Å)/Fe(50 Å) bilayer samples are shown in Figs. 3(a) and 3(b), respectively. The SAED patterns of both samples are shown as insets in these figures. In Fig. 3(a), the HRTEM image of the $t_{Fe}$ = 10 Å sample shows the presence of dispersed nanocrystalline grains (marked by solid lines) embedded in an amorphous matrix. The crystalline regions depicted in the images show regular atomic arrangements. From the interplanar spacings, these regions were identified as the Fe grains. The corresponding SAED pattern (inset) shows only one diffraction ring, which corresponds to the Fe(110) plane. This confirms that the Fe nanocrystals grew with a preferential (110) orientation. The absence of any other diffraction rings suggests that the film is mostly amorphous with a very limited crystallinity fraction. In contrast, the HRTEM image of the $t_{Fe}$ = 50 Å sample (Fig. 3b) exhibits a more distinct fringe pattern, indicating the presence of multiple, large-sized and higher-density grains. The corresponding SAED pattern (inset of Fig.3b) shows multiple rings of Fe grains and two Si rings. This indicates the coexistence of Si and body-centered cubic (bcc) Fe grains of random orientations. The larger number of Fe diffraction rings with enhanced sharpness reflects the higher degree of crystallinity of the sample due to the enhanced grain growth at higher thicknesses. Furthermore, the images also reveal that the separation between the Fe grains is larger for the $t_{Fe}$ = 10 Å sample than for the $t_{Fe}$ = 50 Å sample. This suggests a transition from a more dispersed Fe nanocrystal in an amorphous Si matrix to a denser network of interconnected Fe and Si nanocrystallites. This percolating microstructural connectivity is expected to enhance the spin-dependent scattering process and strongly influence the magnetotransport properties of the films.

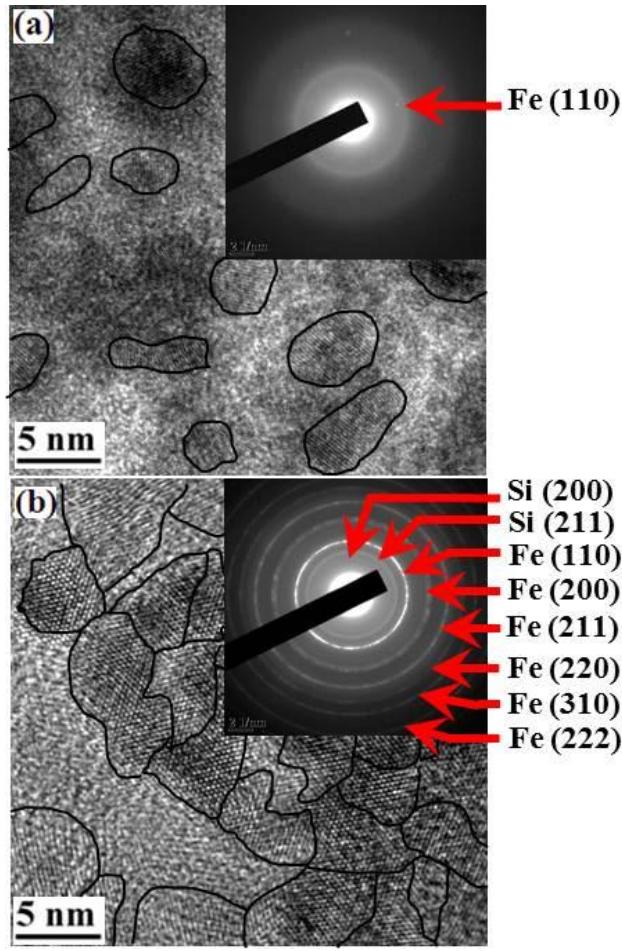

**Fig. 3** (a) HRTEM image of the Si(50 Å)/Fe(10 Å) bilayer. The solid curves are drawn to indicate the boundaries of the Fe grains. The inset shows the corresponding SAED pattern. (b) HRTEM image of the Si(50 Å)/Fe(50 Å) bilayer showing Fe grains that are very close to each other. The inset shows the corresponding SAED pattern.

**(C) Magnetic properties**

Figure 4 shows the in-plane and perpendicular magnetization data of the Si/Fe bilayer samples at room temperature for various $t_{Fe}$ values. The easier saturation of the in-plane M-H loops as compared to the perpendicular ones confirms the presence of in-plane magnetic anisotropy in the samples. Both the in-plane and perpendicular loops become similar for $t_{Fe} <$ 20 Å, suggesting the presence of superparamagnetic Fe grains which do not show magnetic anisotropy.

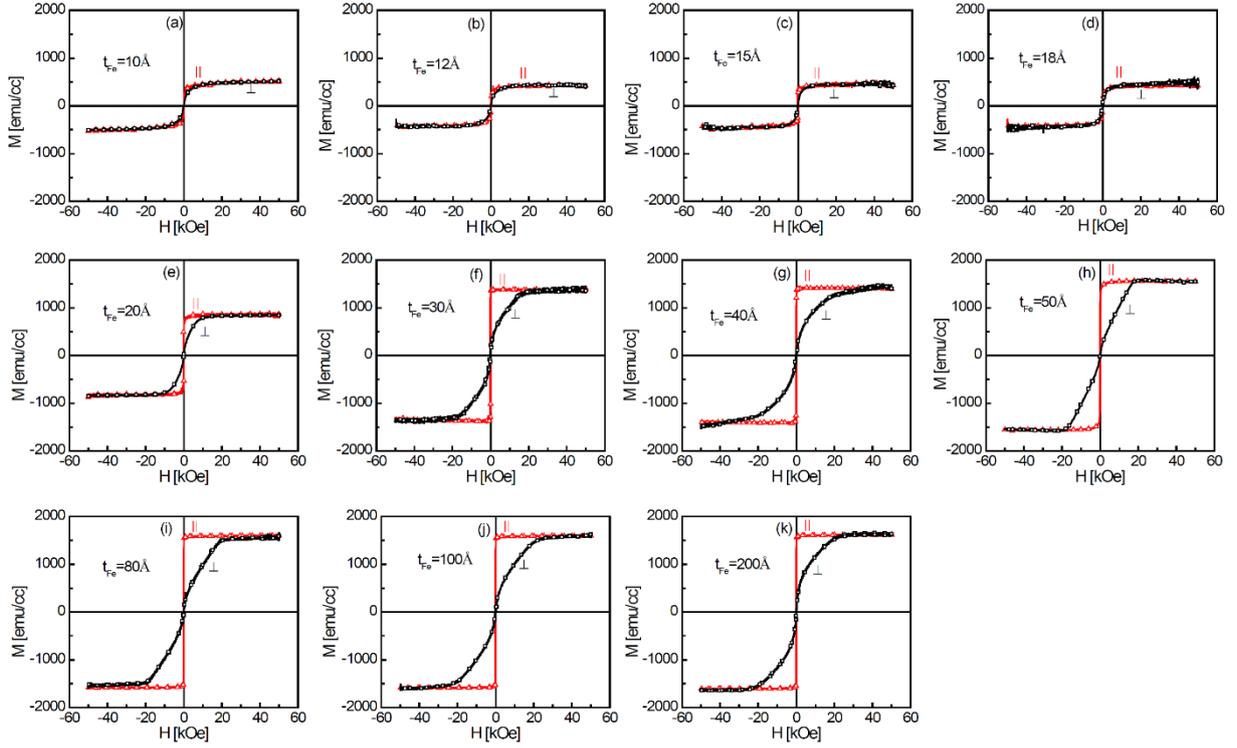

**Fig. 4** Panels (a–k): The in-plane (∥) and perpendicular (⊥) magnetization loops of the Si(50 Å)/Fe($t_{Fe}$) bilayers of different $t_{Fe}$ at 300 K.

From the analysis of the magnetization data, the saturation magnetization $M_s$ is estimated and is plotted as a function of $t_{Fe}$ as shown in Fig. 5(a). The $M_s$ of the samples shows bulk-like saturation around 1550 emu/cc for $t_{Fe} \geq 50$ Å, which decreases sharply to 390 emu/cc when $t_{Fe}$ is decreased to 10 Å. The reported value of $M_s$ for bulk Fe is 1714 emu/cc [26]. A similar decrease in $M_s$ has also been reported for magnetic multilayers with reduced thicknesses [8,27]. In thin films, growth often occurs via island morphology and a weakly connected percolating network before forming a continuous layer. In the discontinuous regime at low thicknesses, the magnetic regions of the film become discontinuous, exhibiting smaller grains, poor crystalline ordering, and strain. These factors contribute to the decrease in $M_s$ by reducing the exchange coupling between the Fe atoms. Besides this, the smaller grains can turn superparamagnetic at room temperature, resulting in a decrease in $M_s$ in the MH loops. As the grains grow above a certain thickness, the film transitions to a continuous and well-coupled ferromagnetic layer, and $M_s$ approaches the bulk value. A decrease in $M_s$ can also be caused by the formation of a magnetic dead layer at the Fe-Si interface due to the interdiffusion between the Fe and Si layers. At smaller thicknesses, this could turn a large portion of the thickness of the film into magnetically inactive, reducing the average $M_s$. The magnetic dead layer thickness ($\Delta$) can be estimated from the $M_s$ vs $t_{Fe}$ data using the equation $M_s(t_{Fe}) = M_s^{bulk}(1 - \Delta/t_{Fe})$ [27]. Fig. 5(b) shows the linear behaviour of the $M_s t_{Fe}$ vs $t_{Fe}$ plot. This results in $M_s^{bulk}$ = 1643 emu/cc and $\Delta$ = 4.7 Å. Note that the fitting was performed using the continuous-regime $M_s$ data (for $t_{Fe} \geq 40$ Å), because reduced thicknesses ($t_{Fe} < 40$ Å) cause large deviations from the straight-line behaviour. This results in an artificially overestimated $\Delta$ value. The larger $M_s$ fitted value reflects the dead layer correction.

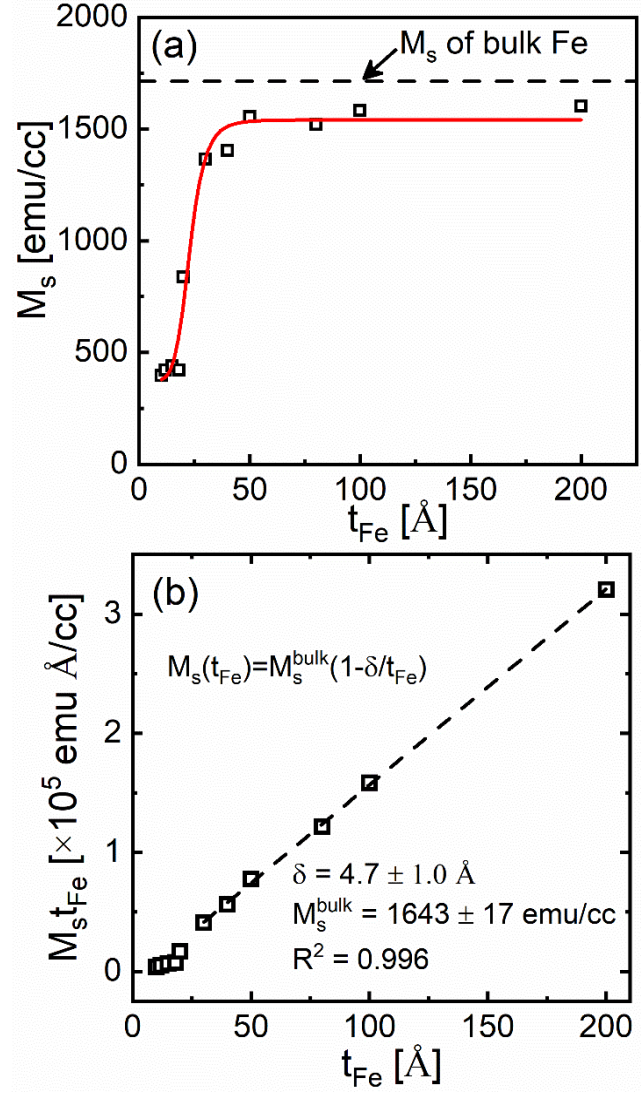

**Fig. 5** (a) The $t_{Fe}$ dependence of $M_s$ for the Si(50 Å)/Fe($t_{Fe}$) bilayers at 300 K, the solid line is a guide to the eye. (b) The $M_s t_{Fe}$ vs $t_{Fe}$ plot for the Si/Fe bilayers. A linear fit using the equation $M_s(t_{Fe}) = M_S^{bulk}\left(1 - \frac{\delta}{t_{Fe}}\right)$, gives $M_S^{bulk} = 1643 \pm 17\ emu/cc$ and $\delta = 4.7 \pm 1.0$ Å.

**(D) Temperature dependence of magnetization**

  The magnetic properties of samples can be strongly influenced by temperature because of the thermal agitation of spins and the changes in the magnetic anisotropy. As the temperature of the sample rises, the spin-spin exchange interaction weakens, resulting in a progressive decrease in magnetization until the sample becomes paramagnetic. A temperature dependence study of magnetisation of the sample reveals information about the stability of the ferromagnetic ordering as the sample undergoes a microstructural transition from continuous to discontinuous state.

We studied the temperature dependence of the magnetization dynamics using zero-field-cooled (ZFC) and field-cooled (FC) measurements. For the ZFC measurements, the sample was first cooled to the lowest temperature 4.2 K, without applying a magnetic field. Then, an in-plane magnetic field of 50 Oe was applied, and the magnetization data were collected while warming the sample from 4.2 to 350 K. For the FC measurements, the sample was cooled in a magnetic field of 50 Oe, and the data were collected while warming up the sample from 4.2 to 350 K. Figures 6(a)–(k) show the ZFC-FC magnetization data of the Si/Fe bilayer samples for different $t_{Fe}$.

The ZFC-FC data clearly demonstrate the evolution of the magnetic properties of the sample with the increase of Fe thickness. The ZFC curves (in red) show a low-temperature peak and then decrease, while the FC curves remain almost flat and exhibit higher magnetization with a slower decrease with temperature. In thin layer ($t_{Fe} \leq 20$Å), panels (a)–(e), the ZFC and FC curves are well separated, with the former showing a low temperature peak (indicative of blocking temperature $T_b$) above which both the curves show a gradual decrease. This suggests the presence of superparamagnetic fine grains, where thermal energy can overcome the magnetic anisotropy. The presence of a broad maximum around the superparamagnetic blocking temperature $T_b$ indicates a size distribution of the Fe grains in these samples. The $T_b$ value in these samples varies from 20 K (for $t_{Fe} = 10$ Å) to 60 K (for $t_{Fe} = 20$ Å).

In the intermediate thickness range (30 Å < $t_{Fe} \leq 100$ Å), the ZFC-FC curves in panels (f)–(j) still differ, but with relatively less pronounced peaks. This is a signature of the onset of long-range ferromagnetic ordering, as the thermal stability of the magnetic domains increases with thickness. The mild decrease of the FC magnetization with temperature in panels (h–j) is expected because the FC data were recorded during warming, and therefore reflect the normal thermal demagnetization of Fe at low field. A stronger magnetic ordering with a shift in the blocking temperature to higher values, evidence of the onset of ferromagnetic-like behaviour. In the thicker layer ($t_{Fe} \geq 200$ Å), panel (k), the ZFC-FC curves almost merge, exhibiting higher magnetization and resembling bulk ferromagnetism in which magnetic domains are stable despite thermal agitation.

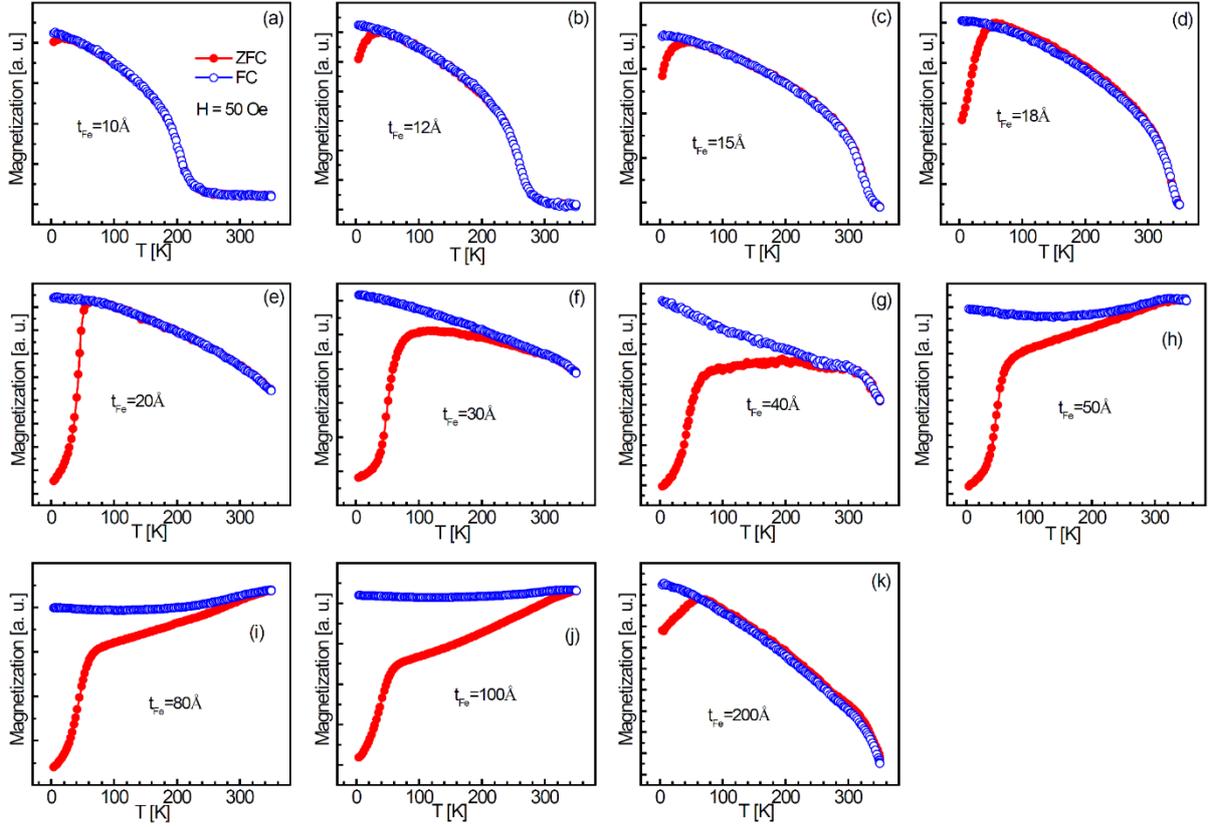

**Fig. 6** (a–k) Temperature dependence of the ZFC-FC magnetization of the Si(50 Å)/Fe($t_{Fe}$) bilayers measured in an in-plane magnetic field of 50 Oe. The ZFC curves (red) were obtained by cooling the sample in zero field and measuring the magnetization during warming in a 50 Oe field. The FC curves (blue) were obtained by cooling the sample in a 50 Oe in-plane field and recording the magnetization during warming in the same field.

**(E) Magnetotransport Properties**

The magnetotransport properties of the Si/Fe bilayer samples were investigated using the anomalous Hall effect (AHE) technique at room temperature. The AHE signal, which is an electrical replica of the magnetization of the sample, is the most convenient method for characterizing the magnetism of nanostructures, where the magnetic signal is too weak to be detected by conventional magnetometry techniques such as PPMS-VSM, SQUID, etc [8]. We adopted the AHE technique for magnetic characterization of our nanostructured samples [7,8,28,29]. The Hall resistivity ($\rho_h$) of a magnetic sample typically consists of two terms: one due to the ordinary Hall effect (OHE) and the other due to the AHE. This is shown by the empirical formula [3,8]:

$$\rho_h = R_H \cdot t_{Fe} = R_0 \cdot B + R_s \cdot 4\pi M$$

Here, $R_H$ is the total Hall resistance, $t_{Fe}$ is the Fe-layer thickness, B is the magnetic field, $R_0$ is the OHE coefficient and $R_s$ is the AHE coefficient, and M is the perpendicular magnetization. The OHE term usually arises due to the Lorentz scattering of the charge carriers, whereas the AHE term has its origin from the breaking of the left-right symmetry during the spin-dependent scattering of the charge carriers at the localized magnetic centers. In the low-field regime, where the magnetization increases rapidly, the total Hall resistivity is dominated by the AHE contribution. However, once the magnetization is saturated, the high-field Hall resistivity is governed by the OHE term, particularly in the thicker Fe films where the linear OHE component ($R_0B$) becomes prominent. The OHE term can be obtained from high-field-regime Hall data where the magnetization saturates. Similarly, the AHE term can be extracted by subtracting the OHE contribution from the total Hall resistivity data, and can be expressed as $\rho_h = R_H \cdot t_{Fe} \sim R_s \cdot 4\pi M$. Fig. 7 shows the typical $\rho_h$ vs H data for Si/Fe bilayers of three different thicknesses 12, 20, and 50 Å, near the percolative regime. These thicknesses correspond to the discontinuous, percolating, and continuous regimes, respectively. The OHE and AHE terms extracted from the Hall data are discussed in the following sections. A reduction in the saturation field is visible with a decrease in $t_{Fe}$.

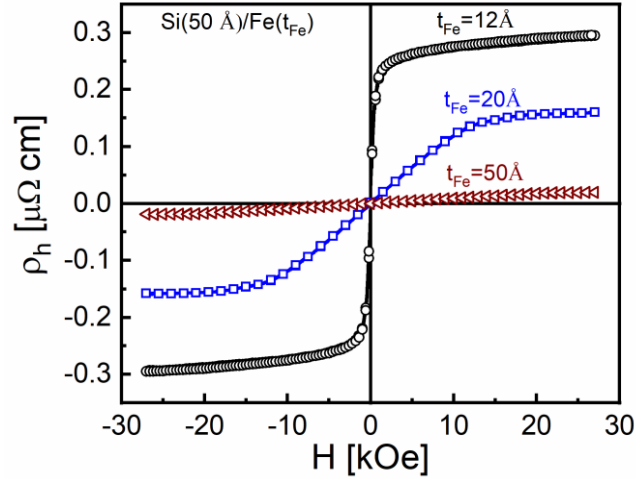

**Fig. 7** Representative Hall resistivity curves of Si(50 Å)/Fe($t_{Fe}$) bilayers for $t_{Fe}$ = 12, 20, and 50 Å near the percolative regime.

### (i) High-Field Hall Response (OHE contribution)

In order to investigate the field dependence of the Hall signal in the Si(50 Å)/Fe($t_{Fe}$) bilayers, we extracted the high-field linear component of the Hall resistivity above magnetization saturation. Since Fe is a multiband ferromagnet, the majority- and minority-spin electrons occupy bands with opposite curvature at the Fermi surface. This results in different Hall responses, with one spin channel producing a negative Hall contribution while the other produces a positive one. The latter dominates due to its larger curvature, leading to an overall positive high-field Hall slope [30,31]. Thus, the high field Hall response is a combined multiband contribution and cannot be interpreted as a pure free electron OHE coefficient. The free-electron expression $R_0 = 1/ne$ is therefore not applicable to ferromagnetic Fe for extracting carrier densities from the OHE data. In ultrathin ferromagnetic films, the high-field Hall response includes contributions from gradual field-induced alignment of spins, as well as anomalous band-structure effects that persist beyond the nominal saturation field. For these reasons, we refer to the extracted slope as an effective high-field coefficient, $R_0^*$. The extracted high-field slope is $1.4 \times 10^{-8}$ Ω·m/T for the 10 Å film, which is nearly three orders of magnitude larger than the value reported for bulk Fe (~ $10^{-11}$ Ω m/T) [32,33]. This reflects the strong inhomogeneity of the ultrathin films and should be interpreted only as an empirical trend.

Figure 8 shows the normalized variation of $R_0^*$ with $t_{Fe}$. In the ultrathin (percolation) regime ($t_{Fe}$ < 40 Å), $R_0^*$ increases rapidly whereas in the thicker regime ($t_{Fe}$ ≥ 40 Å), it shows a decrease towards a nearly constant value. This behaviour reflects the transition from an electronically inhomogeneous film at small thicknesses to a more continuous, bulk-like metallic film at larger thicknesses. The increase in $R_0^*$ in the ultrathin regime is consistent with reduced electronic connectivity, enhanced scattering at Fe-Si interfaces, and finite-size effects, which are commonly observed in the ultrathin regime of ferromagnetic and granular films.

Because the high-field slope has multiple contributions, including OHE and spin-alignment effects, we do not follow quantitative percolation fitting here. Instead, we empirically demonstrate that the thickness evolution of $R_0^*$ is an indicator of the microstructural and magnetic crossover identified independently in the structural (Fig. 2) and magnetization (Fig. 6) measurements.

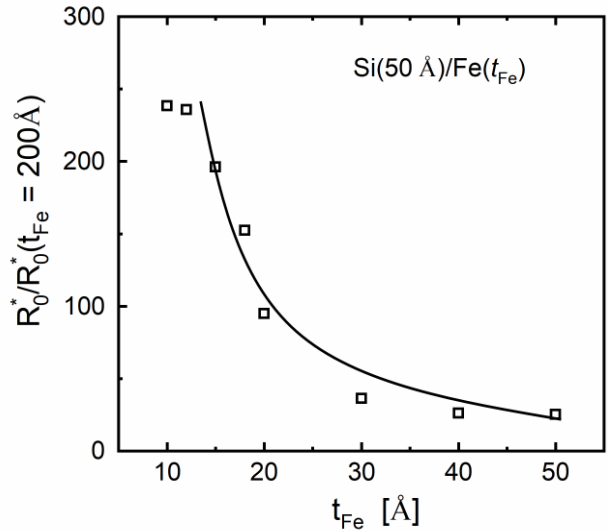

**Fig. 8** Normalized effective high-field Hall slope $R_0^*/R_0^*(t_{Fe} = 200$ Å$)$ of the Si(50 Å)/Fe($t_{Fe}$) bilayers as a function of $t_{Fe}$. The strong enhancement in the ultrathin regime reflects reduced electronic connectivity and surface/interfacial scattering effects. The solid line is a guide to the eye.

**(ii) Anomalous Hall effect (AHE)**

We analyzed the thickness dependence of the anomalous Hall resistivity $\rho_{h,s}^A$ within the framework of the percolation model for a thickness range of 18–50 Å. Fitting the $\rho_{h,s}^A$ data with the equation $\rho_{h,s}^A = A.(t_{Fe} - t_{Fe,c})^{-\mu}$ yields $t_{Fe,c} = 12.0 \pm 0.7$ Å and µ = 1.11 ± 0.05 (Fig. 9). The corresponding log-log scale fit (shown inset) confirms linearity in the choosen range 18 Å ≤ $t_{Fe}$ ≤ 50 Å with µ = 1.16 ± 0.09. The obtained $t_{Fe,c}$ value is consistent with that obtained from the thickness dependence of resistivity data (Fig. 10), while the the critical exponent µ shifts higher to 1.11, suggesting additional influence from magnetic grain connectivity and saturation. The presence of a magnetic dead layer and a nonuniform magnetisation configuration near the percolation threshold could shift the onset of AHE and steepen its thickness dependence. The effective lower cutoff for $t_{Fe}$ in the AHE fit was chosen 18 Å to avoid such magnetic rounding error. The obtained µ=1.11 value is lower than the theoretically predicted value of 1.2-1.3. This suggests the prominence of impurity scattering, the ferromagnetic-semiconductor interface, inhomogeneous magnetic domains, interface effects, roughness, inhomogeneous growth, and finite-size effects in ultra-thin films.

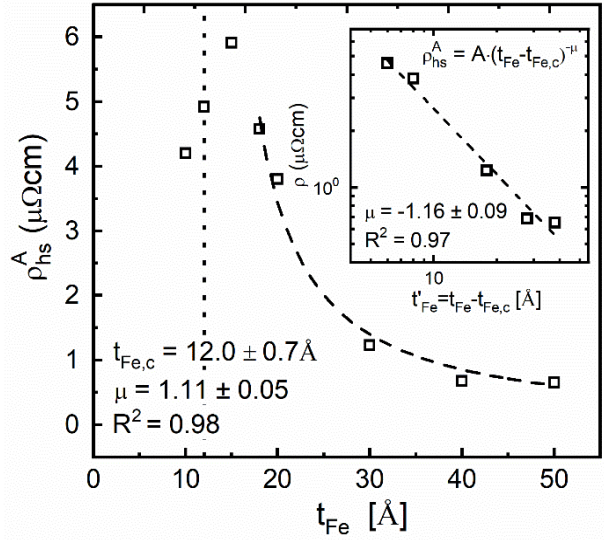

**Fig. 9** The percolation behavior of the $t_{Fe}$-dependence $\rho^A_{h,s}$ of the Si(50 Å)/Fe($t_{Fe}$) bilayers for 18 Å ≤ $t_{Fe}$ ≤ 50 Å. The data were fitted using the percolation law $\rho^A_{h,s} = A \cdot (t_{Fe} - t_{Fe,c})^{-\mu}$, yielding $t_{Fe,c}$ = 12.0 ± 0.7 Å, µ = 1.11 ±.0.05. The inset shows the corresponding log-log scaling plot that confirms the linear behaviour with µ = 1.16 ± 0.09.

### (iii) Percolation analysis of Longitudinal Resistivity (ρ)

The thickness dependence of ρ of the samples was analysed within the same percolative framework and the results are shown in Fig. 10. The nonlinear curve fitting using the percolation law $\rho = A \cdot (t_{Fe} - t_{Fe,c})^{-\mu}$ for thickness range ($t_{Fe}$ = 15–50 Å) yields $t_{Fe,c}$ = 12.1 ± 0.2 Å, µ = 0.87 ± 0.03. The obtained $t_{Fe,c}$ value from fitting is consistent with our earlier observations on $\rho^A_{h,s}$, and indicates a common origin in the percolating transition. The corresponding log-log scale plot, shown in the inset of Fig. 10, reveals the straight line behavior, resulting in µ = 0.78 ± 0.05, which differs slightly from the earlier values. This suggests that ρ follows the divergence near threshold at a relatively slower rate as compared to the combined electronic and magnetic effect as in the previous cases of $\rho^A_{h,s}$.versus $t_{Fe}$. The large deviation of ρ in the percolation regime suggests the scattering of charge carriers due to the microstructural inhomogeneity of grains, scattering at grain boundaries, surface-interface effects, and inhomogeneity of conductivity near the metal-semiconductor interface.

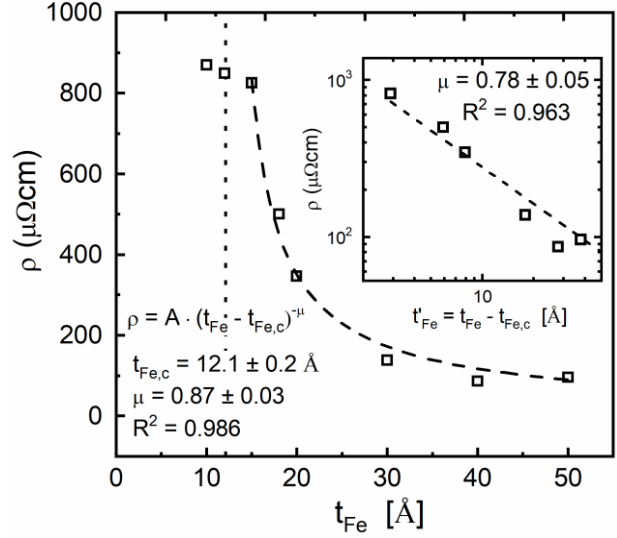

**Fig. 10** Percolation analysis of resistivity ρ vs $t_{Fe}$ plot with $t_{Fe}$ in the range 12–50 Å. The dashed line is the nonlinear fit using the relation $\rho = A \cdot (t_{Fe} - t_{Fe,c})^{-\mu}$ with $t_{Fe,c}$ = 12.1 ± 0.2 Å, μ = 0.87 ± 0.03. The dotted line marks the percolation threshold $t_{Fe,c}$. The inset shows a log-log plot of ρ vs $t'_{Fe}$ (= $t_{Fe}$ - $t_{Fe,c}$) with a slope of 0.78, confirming power-law behavior.

**(iv) Mechanism of AHE**

Since the AHE and longitudinal resistivity have their origin from a scattering event, the two quantities can be scaled using a scaling relation: $\rho^A_{h,s} \propto \rho^n$, where the scaling power n decides the mechanism of AHE whether it is intrinsic or extrinsic (skew scattering/side jump) mechanisms [3,8]. A smaller *n* value, closer to 1, corresponds to skew scattering as the dominant mechanism, whereas a higher n value (~ 2) suggests intrinsic or side jump as the dominant mechanism of AHE. An exponent between 1 and 2 indicates that skew scattering coexists with a quadratic contribution, which may be due to side-jump, intrinsic effects, or both. The intrinsic mechanism of AHE is often referred to as the Berry curvature in momentum space, as is the case in pure bulk systems [3]. In contrast, the extrinsic mechanism is often associated with the asymmetric scattering of spin-polarised electrons with magnetic impurities. A more detailed description of these mechanisms has been provided elsewhere [8].

A comparison of $M_s(t_{\text{Fe}})$ and $\rho_{h,s}^A(t_{\text{Fe}})$ reveals that these two quantities exhibit opposite thickness-dependent trends. The saturation magnetization $M_s$ increases with $t_{\text{Fe}}$ as the film gradually transitions from a percolative network of isolated Fe islands into a more continuous magnetic layer, reducing dead-layer effects, enhancing magnetic connectivity and suppressing spin-wave excitations. In contrast, the anomalous Hall resistivity $\rho_{h,s}^A(t_{\text{Fe}})$, decreases with increasing $t_{\text{Fe}}$ despite the rise in $M_s$. This behaviour follows from the relation $\rho_{h,s}^A(t_{\text{Fe}}) \sim R_s \cdot 4\pi M$, where the $R_s$ scales with the $\rho$ as $R_s \propto \rho^n$, giving $\rho_{h,s}^A \propto \rho^n$. As the Fe film becomes thicker, disorder and grain-boundary scattering are diminshed, leading to a substantial reduction in $\rho$ and hence $\rho_{h,s}^A$ also decreases. At small $t_{\text{Fe}}$, however, enhanced interfacial scattering and roughness increase $\rho$, thereby elevating $\rho_{h,s}^A$ even though $M_s$ is reduced, further strengthening the anticorrelation.

Figure 11 shows the scaling relationship between $\rho_{h,s}^A$ and $\rho$ in a log-log plot. The scaling relation $\rho_{h,s}^A \propto \rho^n$ was evaluated for all films using room-temperature (300 K) transport data. The fitting was therefore performed at a fixed temperature, as our analysis focuses on thickness-dependent scattering mechanisms rather than temperature-dependent scaling behavior. In the high thickness regime ($t_{\text{Fe}} > 30$ Å), the scaling yields n=1.42 ± 0.12, which lies between the skew scattering limit (n = 1) and quadratic (n = 2) behavior expected from intrinsic and side-jump mechanisms. This indicates that the AHE in the low-resistive thick films (red) arises from a mixed contribution of intrinsic and side-jump processes, with skew scattering term becoming comparatively less significant. In contrast, in the thinner ($t_{\text{Fe}} < 30$ Å), high-resistivity films (black), the n = 0.61 ± 0.12, suggesting the $\rho_{h,s}^A$ grows more weakly with $\rho$ in the discontinuous limit with weakly connected grains. This sublinear growth is consistent with skew-leaning behavior modified by strong disorder and interface/grain boundary scattering. These two slope regimes in the scaling plot provides a clear evidence of a crossover in the AHE mechanism, from a mixed intrinsic/side jump dominated response in thicker, continuous films to a disorder-enhanced, skew scattering leaning behaviour in the percolative ultrathin films. A schematic presentation of the morphological transition of the Fe layer from isolated island type ($t_{Fe} < t_{Fe,c}$) to a percolative network ($t_{Fe} \geq t_{Fe,c}$) and finally to a continuous film ($t_{Fe} \gg t_{Fe,c}$), along with the corresponding behaviour the $\rho_h$ and $\rho$ is shown in Fig. 1.

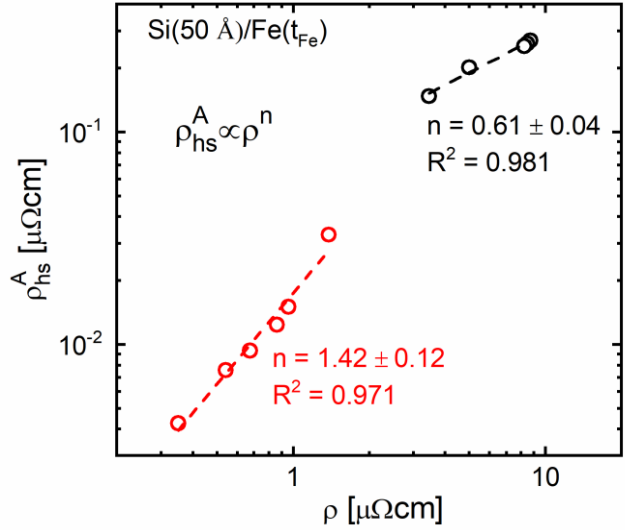

**Fig. 11** The scaling relation between $\rho_{h,s}^{A}$ and $\rho$ of the Si(50Å)/Fe($t_{Fe}$) bilayers was evaluated at room temperature using $\rho_{h,s}^{A} \propto \rho^{n}$. A log-log plot using this equation gives n = 1.42 ± 0.12 for thicker ($t_{Fe} \geq 30$ Å), low resistivity films (shown in red), indicating a mixed intrinsic/side-jump contribution with a small skew-scattering component. In contrast, in the thinner ($t_{Fe} < 30$ Å) high resistive samples, the obtained n = 0.61 ± 0.05, indicating skew scattering dominance in the discontinuous regime.

## 4. Conclusion

In summary, we systematically investigated the effects of percolation-driven transitions on the structural, microstructural, magnetic, and magnetotransport evolutions in Si/Fe bilayers by varying the Fe layer thickness from 10 – 200 Å. By effectively tuning the growth parameters, we encountered three regimes: discontinuous regime (isolated islands), weakly connected percolating network regime, and continuous metallic bulk regime. Structural and microstructural studies revealed a morphological transition from the discontinuous to the continuous regime by increasing the thickness, which is consistent with a percolation-driven growth mode. Magnetic measurements also evidenced a corresponding evolution in magnetic properties: the $M_s$, which exhibits bulk-like saturation above 50 Å, decreases rapidly with the decrease of $t_{Fe}$

Transport measurements, including resistivity, the effective high field coefficient Hall coefficient, and the anomalous Hall resistivity, show critical behavior near the percolation threshold, $t_{Fe,c}$, indicating that the electrical conductivity and the magnetic response can be modified by the connectivity of the electronic and magnetic pathways. The critical exponents from all measurements lie close to the two-dimensional percolation values for disordered metallic systems. AHE scaling reveals a crossover of the AHE mechanism from a mixed intrinsic/side jump contribution in thick, continuous films to a disorder-affected, skew-scattering dominated response in percolative thin films. These results highlight the potential for controlling the AHE response in disordered ferromagnetic films by effectively tuning microstructural connectivity, offering opportunities for spintronics and magnetic sensor applications.

# References

[1]  I. Žutić, J. Fabian, S. Das Sarma, Spintronics: Fundamentals and applications, Rev. Mod. Phys. 76 (2004) 323–410. https://doi.org/10.1103/RevModPhys.76.323.

[2]  C. Chappert, A. Fert, F.N. Van Dau, The emergence of spin electronics in data storage, Nature Mater 6 (2007) 813–823. https://doi.org/10.1038/nmat2024.

[3]  N. Nagaosa, J. Sinova, S. Onoda, A.H. MacDonald, N.P. Ong, Anomalous Hall effect, Rev. Mod. Phys. 82 (2010) 1539–1592. https://doi.org/10.1103/RevModPhys.82.1539.

[4]  A. Fert, H. Jaffrès, Conditions for efficient spin injection from a ferromagnetic metal into a semiconductor, Phys. Rev. B 64 (2001) 184420. https://doi.org/10.1103/PhysRevB.64.184420.

[5]  B.D. Schultz, N. Marom, D. Naveh, X. Lou, C. Adelmann, J. Strand, P.A. Crowell, L. Kronik, C.J. Palmstrøm, Spin injection across the Fe/GaAs interface: Role of interfacial ordering, Phys. Rev. B 80 (2009) 201309. https://doi.org/10.1103/PhysRevB.80.201309.

[6]  R. Jansen, Silicon spintronics, Nat Mater 11 (2012) 400–408. https://doi.org/10.1038/nmat3293.

[7]  S.S. Das, M. Senthil Kumar, Giant anomalous Hall effect in ultrathin Si/Fe bilayers, Materials Letters 142 (2015) 317–319. https://doi.org/10.1016/j.matlet.2014.12.042.

[8]  S.S. Das, M.S. Kumar, Enhancement of anomalous Hall effect in Si/Fe multilayers, J. Phys. D: Appl. Phys. 46 (2013) 375003. https://doi.org/10.1088/0022-3727/46/37/375003.

[9]  G.J. Strijkers, J.T. Kohlhepp, H.J.M. Swagten, W.J.M. de Jonge, Formation of nonmagnetic $c-Fe_{1-x}Si_x$ in antiferromagnetically coupled epitaxial Fe/Si/Fe, Phys. Rev. B 60 (1999) 9583–9587. https://doi.org/10.1103/PhysRevB.60.9583.

[10]  G. Garreau, S. Hajjar, J.L. Bubendorff, C. Pirri, D. Berling, A. Mehdaoui, R. Stephan, P. Wetzel, S. Zabrocki, G. Gewinner, S. Boukari, E. Beaurepaire, Growth and magnetic anisotropy of Fe films deposited on Si(111) using an ultrathin iron silicide template, Phys. Rev. B 71 (2005) 094430. https://doi.org/10.1103/PhysRevB.71.094430.
26